\documentclass[a4paper,fleqn,usenatbib]{mnras}

\usepackage[T1]{fontenc}
\usepackage{ae,aecompl}
\usepackage[dvipsnames]{xcolor}

\usepackage{graphicx}	
\usepackage{epstopdf}
\usepackage{amsmath}	
\usepackage{amssymb}	
\usepackage{listings}

\newcommand{\msun}{$h^{-1}$~M$_\odot$}

\title[A model of the cosmic infrared background]{A Model of the Cosmic Infrared Background Produced
by Distant Galaxies}

\author[S. V. Pilipenko et al.]{
S. V. Pilipenko,$^{1}$\thanks{spilipenko@asc.rssi.ru (SVP)}
M. V. Tkachev$^{1}$,
A. A. Ermash$^{1}$,
T. I. Larchenkova$^{1}$,
\newauthor
E. V. Mikheeva$^{1}$,
V. N. Lukash$^{1}$
\\
$^{1}$Astro Space Center, P. N. Lebedev Physical Institute of RAS, Profsojuznaya 84/32, Moscow 117997, Russia\\
}

\date{Accepted XXX. Received YYY; in original form ZZZ}

\pubyear{2017}

\begin{document}
\label{firstpage}
\pagerange{\pageref{firstpage}--\pageref{lastpage}}
\maketitle

\begin{abstract}
The extragalactic background radiation produced by distant galaxies emitting in the far infrared
limits the sensitivity of telescopes operating in this range due to confusion. We have constructed a model
of the infrared background based on numerical simulations of the large-scale structure of the Universe
and the evolution of dark matter halos. The predictions of this model agree well with the existing data on
source counts. We have constructed maps of a sky field with an area of 1\,deg$^2$ directly from our simulated
observations and measured the confusion limit. At wavelengths $100-300$\,$\mu$m the confusion limit for a
10-m telescope has been shown to be at least an order of magnitude lower than that for a 3.5-m one. A
spectral analysis of the simulated infrared background maps clearly reveals the large-scale structure of the
Universe. The two-dimensional power spectrum of these maps has turned out to be close to that measured
by space observatories in the infrared. However, the fluctuations in the number of intensity peaks observed
in the simulated field show no clear correlation with superclusters of galaxies; the large-scale structure has
virtually no effect on the confusion limit.
\end{abstract}



\section{Introduction}
The sky background consists of many components
that can be divided into two groups. The first group
includes all of the components associated with our
Galaxy and the Solar system, such as the zodiacal
light and the Galactic ``cirrus''. The second group is
not associated with our Galaxy. The spectrum of this
component has two peaks. The first peak produced by
the cosmic microwave background (CMB) is in the
millimeter wavelength range, while the second one is
in the far infrared (IR). The latter is called the cosmic
IR background (CIB). The CIB in the wavelength
range 100--600\,$\mu$m is believed to be due mainly to
the emission of galaxies at redshifts $z \sim 1-2$ with
active star formation and a large amount of dust that
reradiates much of the starlight in the far IR with an
intensity maximum near 100\,$\mu$m (in the rest frame).

At an angular resolution that the single telescopes
designed for far-IR observations usually have, the
CIB is not resolved completely into individual sources
and consists of spots with different brightnesses. The
spatial CIB fluctuations create the so-called confusion 
problem, where the faint point sources cannot be
separated from the spots produced by many distant
galaxies. This effect limits the sensitivity of photometric 
studies in the far-IR and submillimeter ranges.

Future far-IR observatories, such as Millimetron
(http://millimetron.ru, \citet{2009ExA....23..221W, 2012SPIE.8442E..4CS, 2014PhyU...57.1199K}),
CALLISTO, and
OST, will be able to distinguish much more details on
IR sky maps than are accessible in the currently available 
observations. Therefore, to predict the influence
of confusion for future observations, it is necessary
to construct a model capable of extrapolating the
current views of the CIB to a higher resolution and
sensitivity.

The goal of this paper is to construct a CIB model
to predict the possibilities of observations with the
Millimetron space telescope and other similar space
observatories. It should be emphasized that this CIB
model is based on the halo catalog produced with one
of the advanced numerical models for the evolution of
the matter distribution in the Universe. This allows
us to study in detail the manifestations of the large-scale structure of the Universe on the CIB maps.
Since a large number of CIB models and approaches
to their construction exist at present, it turns out to
be important to classify these models and approaches
as well as to choose the most suitable ones for our
purposes from them.

\section{Classification of CIB models}
o construct a CIB model, it is necessary to specify
the spatial distribution of galaxies and their spectra.
The existing approaches to modeling the CIB can
be divides into three groups: backward evolution,
forward evolution, and a semi-analytical method. The
first two approaches (methods) are described in the
review by \citet{1996AIPC..348..147L}. Let us briefly consider the
approaches listed above.

The method of backward evolution is based on the
assumption that the evolution of model parameters
known from observations of the local Universe and
observations at some redshifts can be specified in the
direction of increasing redshift, i.e., in the direction
opposite to the natural evolution (from the formation
of objects to the present day). Many papers are de-
voted to modeling by this method (see, e.g. \citet{2010A&A...517A..74F,2009MNRAS.394..117R,2009ApJ...701.1814V,2003MNRAS.338..555L,2003ApJ...585..617D,2011A&A...529A...4B}). Usually, a known luminosity function is taken in the method of backward
evolution and its parameters are assumed to evolve
with redshift z. The free model parameters determine
precisely how they evolve. A library of spectra is used
to describe several galaxy populations. Depending on
model complexity, the shapes of the spectra may or
may not depend on z or luminosity. For example,
there can be two \citep{2011A&A...529A...4B,2012ApJ...757L..23B,2003ApJ...585..617D}, four \citep{2006MNRAS.369..281J,2009MNRAS.394..117R} or five \citep{2010A&A...517A..74F,2011MNRAS.416...70G} galaxy populations. In some
papers, instead of approximating the evolution of the
luminosity function, it is calculated from the evolution
of the star formation rate \citep{2012ApJ...757L..23B}
or
from the assumption that the stellar mass function
evolves with star formation and other processes
are taken into account \citep{2012MNRAS.426.2797W}.
It is possible to use the color–luminosity function
rather than the luminosity function \citep{2011MNRAS.418..176R}.

Different approaches are also possible in the methods
 of determining the free parameters. For example,
all of the available source counts and/or luminosity 
functions \citep{2011A&A...529A...4B,2010A&A...517A..74F,2011MNRAS.416...70G,2011MNRAS.418..176R,2009MNRAS.394..117R,2009ApJ...701.1814V}, CIB spectra \citep{2010A&A...517A..74F}, redshift distributions of objects with fluxes above a
certain one for different wavelengths \citep{2010A&A...517A..74F,2011MNRAS.418..176R} can be
used, or the parameters are specified from model assumptions \citep{2012ApJ...757L..23B,2003ApJ...585..617D}.

The effects of strong gravitational lensing of
sources at high redshifts should also be taken into
account. The need for this is confirmed by the
detection of bright lensed objects at high redshifts
in the (sub)millimeter range \citep{2013Natur.495..344V,2010Sci...330..800N}.
Both a magnification of the
source and a change in its angular size must be
observed in the case of strong gravitational lensing.
Its multiple images can also be observed in the case
of a sufficient telescope resolution. Since the submillimeter 
telescopes have a limited spatial resolution
and, consequently, are subjected to confusion, the
lensed unresolved distant sources must contribute
significantly to the background radiation at submillimeter 
wavelengths. The expected number of lensed
sources at 500\,$\mu$m with a flux of more than 100\,mJy
is about 15\%, and this fraction increases to 40\% at
about 1\,mm \citep{2011A&A...529A...4B}.

It should be noted that the method of backward
evolution is not based on any physical model; it is
based largely on observational data fitting. If we
choose the shape of the luminosity function, choose
the spectra, and carefully specify or fit the free parameters, 
then any observational data can be fitted
in principle. Thus, this method has a good descriptive 
power but a very poor predictive one. At a
small amount of observational data its application can
lead to results far from reality (see, e.g., \citet{2011MNRAS.410.2556D}).

It is important to note that only the source counts
as a function of the flux, redshift, and luminosity are
obtained at the ``output'' of the method of backward
evolution. This approach gives no information about
the spatial distribution of sources and the association
of galaxies with the distribution of dark matter.

The method of forward evolution is also described
in detail in many papers (see, e.g., \citet{1996AIPC..348..147L,2000A&A...363..851D,2010ApJ...718.1001B}).
Its basic idea is the construction of a semi-analytical
model for the evolution of dark matter halos in
accordance, for example, with the Press–Schechter
formalism. The halo mass function and its evolution 
with redshift are calculated from this model.
Thereafter, the mass function is converted to the
luminosity functions of one or more types of galaxies
by taking into account their evolution \citep{2009ApJ...696..620C,2008ApJS..175..356H},
the star formation processes, the evolution of stars, the feedback,
dust absorption and emission effects \citep{1998MNRAS.295..877G,2000A&A...363..851D,2010ApJ...718.1001B,2004ApJ...600..580G,2010MNRAS.405....2L,2000MNRAS.319..168C}, the accretion onto black hole \citep{2004ApJ...600..580G,2008ApJS..175..356H}, etc.

A major shortcoming of the classical approach of
forward evolution is that the results obtained characterize the background only on average. This method,
along with the method of backward evolution, makes
it possible to determine the number of sources in
a specified range of wavelengths and fluxes but not
properties (coordinates, luminosities, spectra) of individual sources.

The third approach, called semi-analytical, was
used in a number of papers to determine the properties
of galaxies in the optical and near-IR ranges (see,
e.g., \citet{2005MNRAS.364..407C,2006MNRAS.370.1651C,2015A&A...575A..32C}).
This approach uses a numerical model for the evolution of dark matter based on which a halo catalog
is produced. Baryonic matter is then added to each
halo. Various processes, such as the accretion of gas
from the parent halo, the star formation, the activity in
galactic nuclei, the feedback, the gas outflow, the formation of a disk, a bulge, etc., are taken into account
in this case. The specific ways of allowance for these
processes are highly varied. In contrast to the method
of forward evolution, in the semi-analytical method
numerical models are used instead of the evolution
of the mass functions to obtain information about the
evolution of dark matter halos.

Various approaches, for example, the iterative
technique, as in \citet{2013MNRAS.431.3373H}, can be
used to specify the free parameters of semi-analytical
models just as in other methods. In recent years
the semi-analytical methods based on numerical
simulations of the dark matter distribution have
gained a major development. In particular, they allow
one to trace a detailed picture of galaxy evolution
and, as a consequence, to obtain the mass functions
of various components, the stellar mass-halo mass
relationship at various $z$, the luminosity functions,
the color--magnitude diagrams, the Tully--Fisher
relations, the color--mass diagram, the evolution of
the star formation rate with redshift, the distribution
of star formation rates in galaxies at various z, the
dependence of the stellar mass of the bulge on the
total mass of the bulge, the evolution of the star density
$\rho_*$\.[$M_\odot$Mpc$^{-1}$] with $z$, the correlation between
the masses of black holes and bulges, the luminosity
function of quasars, the evolution of the fraction of
active galactic nuclei (AGNs), and the evolution of
the black hole mass function.

It should be noted that the approach used here can
be attributed to this type, i.e., to the semi-analytical
approach.

\section{Model description}
\subsection{Model requirements}
The goal of model construction is to determine the
possibilities of observing objects with low fluxes (faint
objects) and to investigate the large-scale structure
of the Universe with space observatories operating in
the far IR and having an aperture of about 10 m. The
following parameters of the Millimetron observatory 
\citep{2012SPIE.8442E..4CS} were taken as a basis:

\begin{enumerate}
\item The angular resolution of the telescope is limited by diffraction at wavelengths $\lambda > 300$\,$\mu$m and is
at least 5\,arcsec at shorter wavelengths. However, the
resolution will not be better than 2\,arcsec even for the
diffraction-limited resolution at 100\,$\mu$m. This means
that galaxies at redshifts $z > 0.5$ may be deemed
point sources.
\item The sensitivity of broadband photometry can
reach 0.1\,$\mu$Jy at wavelengths 100--400\,$\mu$m, and the
simulated catalog of galaxies must be statistically
complete to such fluxes.
\item The field of view of the telescope is about
6\,arcmin; therefore, simulated survey fields of several
square degrees must be sufficient to simulate the
observations.
\item The model should take into account the nonlinear pattern of the large-scale matter distribution in
the Universe, in particular, the presence of clusters
and superclusters of galaxies that can affect the observed background characteristics (both directly and
through gravitational lensing).
\end{enumerate}

From the above-listed requirements we constructed a semi-analytical model based on cosmological numerical simulations. Its construction
consisted of the following main steps:

\begin{enumerate}
\item Presenting the halo catalog in the form of a
filled cone.
\item Assigning the luminosity to each halo.
\item Calculating the fluxes from each galaxy for a
specified wavelength (spectral window).
\item Mapping the sky using ``cloud-in-cell'' interpolation.
\item Convolving the map with the telescope beam.
\end{enumerate}

The main difference between our model and those
existing in the literature for the far IR is allowance for
the nonlinear large-scale structure of the Universe.
Below we will consider the listed steps in more detail.

\subsection{The Cone, the Halo Properties, and the Large-Scale Structure}

To produce the halo catalog, we use the COSMOSIM database, from which we extracted all the
available cuts in time of the Small MultiDark Planck
numerical model \citep{2016MNRAS.457.4340K} with a cube size
of 400\,Mpc/$h$ (where $h = H_0/100$, $H_0$ is the current
Hubble constant). The cone is $1^\circ \times 1^\circ$ in size, corresponding 
to a size of its base $\sim$ 100\,Mpc/$h$ at $z = 6.2$.
The cone axis crosses the cube of the numerical model
at such an angle that the same part of the cube falls
into the cone only once. Depending on the distance
to the cone vertex, we determined the closest cut in
time from which the halo coordinate were taken. This
ensured the continuity of the elements of the large-
scale structure along the cone (Fig. 1).

We checked that increasing the minimum halo
mass to $10^{11}$\,$M_\odot$ or decreasing the maximum halo
mass to $10^{14}$\,$M_\odot$ did not affect the results within the
accuracy of our method. The maximum redshift does
not affects the results either. Changing the minimum
redshift affects noticeably the source counts at the
shortest wavelengths 70--100\,$\mu$m. The best correspondence
to the known source counts is observed at
$z = 0.54$ (Fig. 2).

\begin{figure*}
\includegraphics[width=\linewidth]{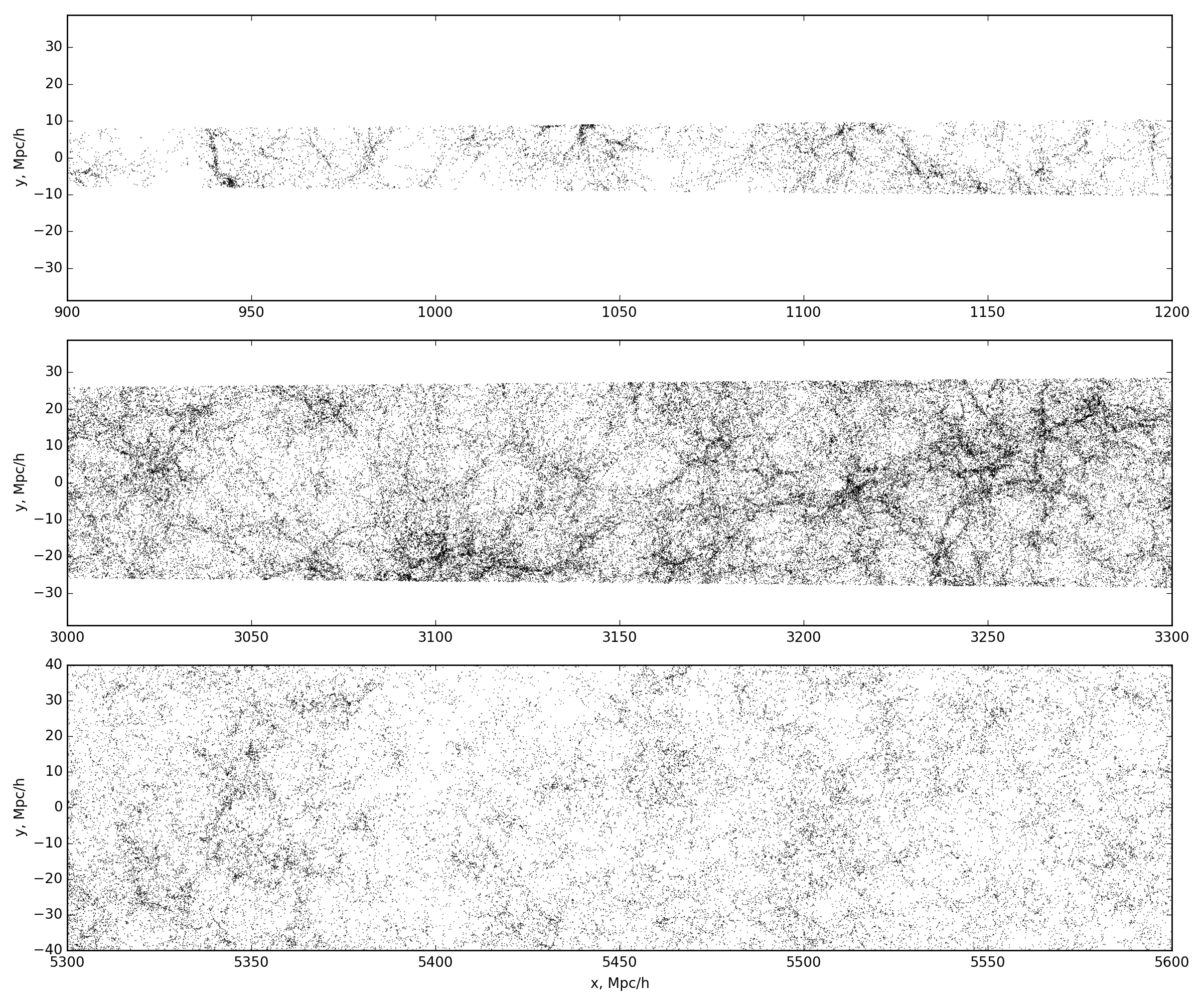}
\caption{Several regions of the cone projection onto the XY plane in comoving coordinates. The X axis is along the cone axis.
Each point is a separate halo.}
\label{fig:proj}
\end{figure*}

\begin{table}
\caption{Parameters of the simulated survey cone}
\begin{center}
\begin{tabular}{lc}
\hline
Parameter & Value \\ \hline
Minimum redshift & 0.30 \\
Maximum redshift & 6.19 \\
Minimum mass & $3\times10^{10}$ \msun \\
Maximum mass & $2.56\times10^{14}$ \msun \\
Number of halos & 1285307 \\
Number of halos $M>10^{13}$ \msun & 936 \\
Number of halos $M>10^{14}$ \msun & 15 \\
\hline
\end{tabular}
\end{center}
\label{tab:cone}
\end{table}

The cone parameters are presented in Table 1.
The cone contains 15 clusters of galaxies and $\sim$ 900
groups. To analyze the large-scale structure in this
cone, we used the minimal spanning tree technique,
which is widely applied in searching for superclusters
\citep{2004A&A...418....7D,2007ARep...51..820P,2013MNRAS.429.2910C}. 
In this case, two parameters are specified:
the maximum threshold branch length needed to include the branch in a cluster and the minimum number of objects in the cluster. Both parameters were
varied. For example, the histogram of the distribution
of tree branches in length has a maximum at a length
of 0.5\,Mpc/$h$. If we take this value as the threshold length and choose 100 objects as the threshold
number, then in our catalog the algorithm detects
11 clusters whose size does not exceed 20\,Mpc/$h$.

\subsection{Gravitaional lensing}

To take into account the gravitational lensing, we
search for all possible pairs of close (in angular separation) halos in the produced catalog and calculate
the magnifications of the more distant halo (source) in
this pair from the observer when lensed by the closer
halo (lens). The magnifications μ were calculated
for two simple lens models: a point lens (PL) and
a singular isothermal sphere (SIS). The expression
for the magnification is known (see, e.g., \citet{1992Schnider}) to be

\begin{equation}
\mu(M_L,D_L,D_S,\beta) = {\tilde{\beta}^2+2 \over \tilde{\beta}\sqrt{\tilde{\beta}^2+4}}
\end{equation}

for the PL model and

\begin{equation}
\mu(M_L,D_L,D_S,\beta) = \begin{cases} {2 \over \tilde{\beta}}, & \tilde{\beta} \leqslant 1 \\ {\tilde{\beta}+1 \over \tilde{\beta}}, & \tilde{\beta}>1 \end{cases},
\end{equation}

for the SIS model, where

\begin{equation}
\tilde{\beta} = {\beta \over \alpha_0},
\end{equation}
$\beta$ is the angular separation between the source and
the lens, $\alpha_0$ is the characteristic angular separation
dependent on the source redshift $z_S$ and the lens
redshift $z_L$.

The characteristic angular separation $\alpha_0$ is written as

\begin{equation}
\alpha_0 = \sqrt{{4G M_L \over c^2}{D_{LS} \over D_S D_L (1+z_L)}},
\end{equation}
where $D_L$ and $D_S$ are the comoving distances from
the observer to the lens and the source, respectively,
$D_{LS}$ is the comoving distance between the lens and
the source, $z_L$ is the lens redshift, $c$ is the speed of
light, and $G$ is the gravitational constant \citep{1992Schnider}.

For our purposes, we took into account the strong
gravitational lensing events with a magnification $\geq2$,
which subsequently allowed us to properly compare
our results for the source counts with those from
\citet{2011A&A...529A...4B}. Furthermore, we took into
account the extent of the emission sources. The
sources were assumed to be circles with a uniform
surface brightness with a radius $\rho$ equal to 0.25\,arcsec. In this case, the maximum magnification for the
PL model is \citep{1987A&A...179...71S}

\begin{equation}
\mu_{max}=\frac{\sqrt{\gamma^2 + 4}}{\gamma},\;\;\gamma = \frac{\rho}{\alpha_0},
\end{equation}

As our calculations show (see Fig. 2), allowance
for the strong gravitational lensing events with a
magnification $\geq$2 in the SIS model turns out to be
important when counting the sources at fluxes larger
than 10\,mJy for wavelengths of 70, 250, and 500\,$\mu$m
and larger than 1\,mJy at a wavelength of 1.2\,mm.
In Fig. 2 the solid line and, for comparison, the dotted line indicate, respectively, the results with and
without gravitational lensing. The PL model gives
slightly poorer agreement with the source counts.
In the model of backward evolution from \citet{2011A&A...529A...4B}, the lensing was also taken into account
and similar results were obtained.

\subsection{Determining the luminosities of galaxies}

In this paper we used one of the most commonly
applied relations between halo mass $M$ and galaxy
luminosity $L$ from \cite{2013ApJ...779...32V}:

\begin{figure*}
\includegraphics[width=\linewidth]{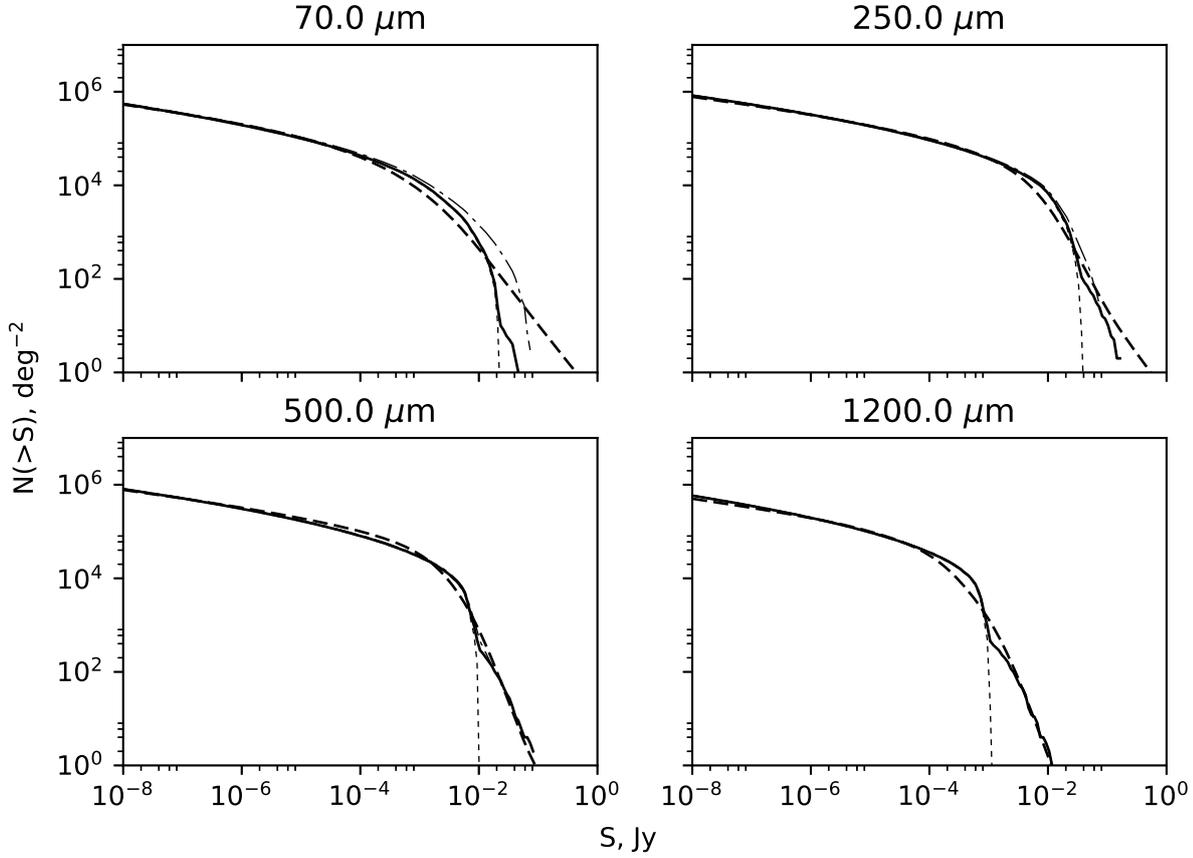}
\caption{Source counts in our model with gravitational lensing (solid lines), without gravitational lensing (dotted line), and in the model of \citet{2011A&A...529A...4B} (dashed line) at 70, 250, 500, and 1200\,$\mu$m. The thin dash--dotted line is our model for the case of minimum redshift $z=0.3$.}
\label{fig:bethermin}
\end{figure*}

\begin{equation}
L(M,z) = L_0 (1+z)^\eta \log(M) \exp \left( -{(\log(M)-\log(M_0))^2 \over 2\sigma_L^2} \right),
\end{equation}
where $\log(M_0) = 12.6$, $\sigma_L^2 = 0.15$, $\eta = 3.16$ at $z<2$ and $\eta = 0$ at $z>2$, $L_0 = 5\times10^9$ L$_\odot$ for the integrated luminosity in the range from 8 to 1000\,$\mu$m
in the galaxy frame. This mass--luminosity relation
is based on the fact that the star formation is most
efficient in a limited range of halo masses, near $M_0$ \citep{Shang12}. 
For masses much higher or
lower than this one the star formation is suppressed
by the feedback effects: photoionization, supernova
explosions, AGN activity \citep{Keres05,Bower06,2006MNRAS.365...11C}.
The model parameters $L_0$, $M_0$, and $\sigma_L$ were determined by fitting the source counts to the results from
\citet{2011A&A...529A...4B}, as the best of the existing fits of the
source counts in different ranges. Note that the
model of backward evolution from this paper is an
updated version of the earlier model proposed in \citet{2003MNRAS.338..555L}.

To determine the model parameters $L_0$, $M_0$ and $\sigma_L$, we used the source counts as a function of the
flux at $\lambda=$ 70, 110, 160, 250, 350, 500, 850, 1200,
and 2000 $\mu$m. The parameters were determined by
minimizing the rms deviation of the source counts in
the same flux intervals as in the data from~\cite{2011A&A...529A...4B}.
The integrated counts were compared for the most accurate reproduction of the faint source
counts.

For each wavelength we calculated the monochromatic flux, i.e., no information about the passband
was used. Note that the data given in~\citet{2011A&A...529A...4B} were obtained with real telescopes in
finite-width passbands. Fitting these data without
allowance for the shape of the passband automatically corrects the parameters of our model in such
a way that the results obtained in it are already not
monochromatic but those obtained with an averaged
passband over all the instruments whose data were
used in \citet{2011A&A...529A...4B}.

The source counts in our model and in the models of \citet{2011A&A...529A...4B} are shown in Fig. 2.
As can be seen from this figure, good agreement of
the results obtained in both models is observed at
longer wavelengths, while a discrepancy is observed
for fluxes 10--100 mJy at short wavelengths.

\subsection{The Spectra of Galaxies}
The shape of the spectral energy distribution was
taken from \citet{2010A&A...514A..67M}. The luminosity
determined by the method described above specifies
its amplitude for each galaxy. Using only one spectrum for all galaxies is a serious simplification of the
model. In the approach adopted here each galaxy has
its unique halo, and one can devise a large number of
ways of specifying the galaxy type as a function of the
formation history of this halo, its spatial environment,
angular momentum, etc. Analysis of this ways is beyond the scope of this paper. Nevertheless, for comparison we also used several other galaxy spectra that
were taken from the library of spectra in \citet{2001ApJ...556..562C}, where 105 model spectra are presented.
We chose the two extreme spectra, with the largest
and smallest fractions of far-IR emission, whereupon
the source count fitting procedure was repeated. As
a result, the counts changed by no more than 30\%,
which is comparable to the accuracy of our model.

A consequence of applying a single spectrum for
all types of galaxies is a slight spread in galaxy luminosity at fixed mass. This spread is caused by
the redshift dependence of the luminosity. For example, galaxies with halo masses of $10^{13}$ M$_\odot$ and
$10^{11}$ M$_\odot$ have IR luminosities from $10^{10}$ to $10^{11}$ L$_\odot$
and from $10^{6}$ to $10^{7}$ L$_\odot$, respectively. The inferred
spread is several times smaller than that observed, for
example, in SDSS galaxies at low redshifts \citep{2004MNRAS.351.1151B}.

\subsection{Comparison with the ALMA Observations}

Recent deep observations at the ALMA observatory \citep{2015A&A...584A..78C,2016ApJS..222....1F}
have yielded for the first time the source counts at
wavelengths 1.1--1.3 mm with fluxes 20--100 $\mu$Jy.
We compared these counts with the predictions of our
model; the results are presented in Table 2. As can be
seen from this table, the predictions of our model are
in good agreement (given the measurement errors)
with the ALMA observations.

\begin{table*}
\caption{Comparison with the ALMA source counts}
\begin{tabular}{lcc}
\hline
Parameter & Observations & Our model   \\ \hline
$\lambda=1.1$ mm {\citep{2015A&A...584A..78C}} & & \\
Integrated flux at $S>130$ $\mu$Jy & $>17.2$ Jy deg$^{-2}$ & 14.8 Jy deg$^{-2}$ \\
$dN/d(\log_{10}S)$ at $S=130$ $\mu$Jy &  $10^{+7}_{-4}\times10^4$ deg$^{-2}$ & $3.5\times10^4$ deg$^{-2}$ \\ \hline
$\lambda=1.3$ mm {\citep{2015A&A...584A..78C}} \\
Integrated flux at $S>60$ $\mu$Jy & $>12.9$ Jy deg$^{-2}$ & 9.5 Jy deg$^{-2}$ \\
$dN/d(\log_{10}S)$ at $S=60$ $\mu$Jy &  $11^{+14}_{-7}\times10^4$ deg$^{-2}$ & $3.8\times10^4$ deg$^{-2}$ \\ \hline
$\lambda=1.2$ mm {\citep{2016ApJS..222....1F}} \\
Integrated flux at $S>20$ $\mu$Jy & $22.9^{+6.7}_{-5.6}$ Jy deg$^{-2}$ & 13.1 Jy deg$^{-2}$ \\
$dN/d(\log_{10}S)$ at $S=20$ $\mu$Jy &  $3.2\times10^5$ deg$^{-2}$ & $5.9\times10^4$ deg$^{-2}$ \\
\hline
\end{tabular}
\label{tab:alma}
\end{table*}

\section{The Identification of Point Sources and the Confusion Limit}

There exist at least two criteria for determining the
confusion limit \citep{2003ApJ...585..617D}: the photometric
and source density criteria. The photometric criterion
is applied to sources that are too faint to be detected
separately. In contrast, the source density criterion
is applied to sources that can be detected as individual objects. Given a simulated sky map, it is also
possible to study the completeness of the catalog of
point sources extracted from this map. This method
is closest to real measurements. Nevertheless, the
result can depend on the applied source identification
algorithm.

In this paper the confusion limit was determined
using the following four methods.

(1) The source density criterion. For sources with
a flux above some threshold value the probability of
having two sources in the telescope beam is less than
10\% \citep{2003ApJ...585..617D}:
\begin{equation}
N_\mathrm{SDC}(>S_\mathrm{lim}) = - {\log (1-P) \over \pi k^2 \theta_{FW}^2},
\end{equation}
where $N_\mathrm{SDC}(>S_\mathrm{lim})$ is the density of sources with a
flux above the threshold value $>S_\mathrm{lim}$, $\theta_{FW}$ is the full
width at half maximum of the beam profile, $P$ is the
probability that the two sources will be indistinguishable, i.e., will be at a distance less than $k\theta_{FW}$, where
$k = 0.8$. As can be seen, only the information about
the source counts is sufficient for this criterion to be
applied.

(2) The photometric criterion:
\begin{equation}
S_\mathrm{lim} = q \sigma(S_\mathrm{lim}),
\label{eq:conf_ph}
\end{equation}
where $\sigma(S_\mathrm{lim})$ is the level of flux fluctuations in the
telescope beam from sources with a flux below the
threshold value $S_\mathrm{lim}$, $q=5$ is the signal-to-noise ratio
needed for successful detection. The value of $\sigma(S_\mathrm{lim})$
can be found in two ways: from the source counts
by assuming their uniform random distribution and
directly from the simulated map.

(3) The photometric criterion with the determination of $\sigma(S_\mathrm{lim})$ from a simulated map containing no
sources brighter than $S_\mathrm{lim}$.

(4) Investigation of the map completeness. It
consists in determining the flux at which the fraction
of the sources found among those initially included in
the model will be 50\%. The search for sources on
the maps was made by several methods. The first
of them is to determine the local maxima using the
second derivatives (we used the code provided to us
by D.I. Novikov). This method yields good results
for a noiseless map, but it cannot be applied for real
astronomical images. The result of the identification
of point sources by this method is shown in Fig. 3.

We used two more methods widely applied in
practice. One of them, SEXTRACTOR \citep{1996A&AS..117..393B}, found much fewer sources than did
the method of local maxima; therefore, we abandoned
this method. The second one is the getsources code
applied to process the images from the Herschel
space observatory \citep{Menshchikov}. This
code has two modes: the monochromatic one, where
the map only at one wavelength is used, and the multiwavelength one. In the latter the map with the best
angular resolution (usually in the shortest wavelength
range) is used to determine the coordinates of sources
on the remaining maps. Getsources showed excellent
results comparable to those of the method of local
maxima in the monochromatic mode, while in the
multiwavelength mode it surpassed the method of
maxima.

\begin{figure*}
\includegraphics[width=0.49\linewidth]{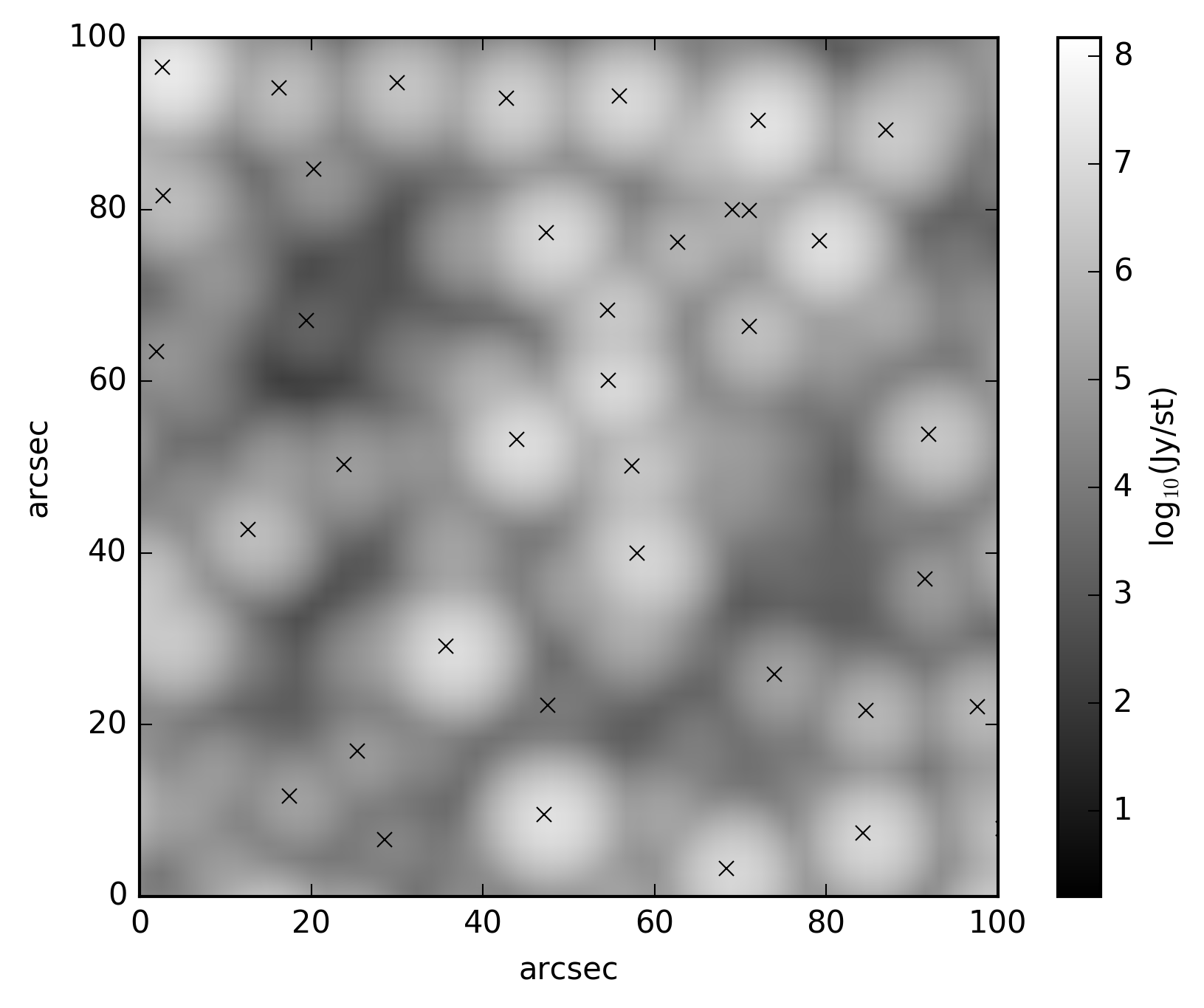}
\includegraphics[width=0.49\linewidth]{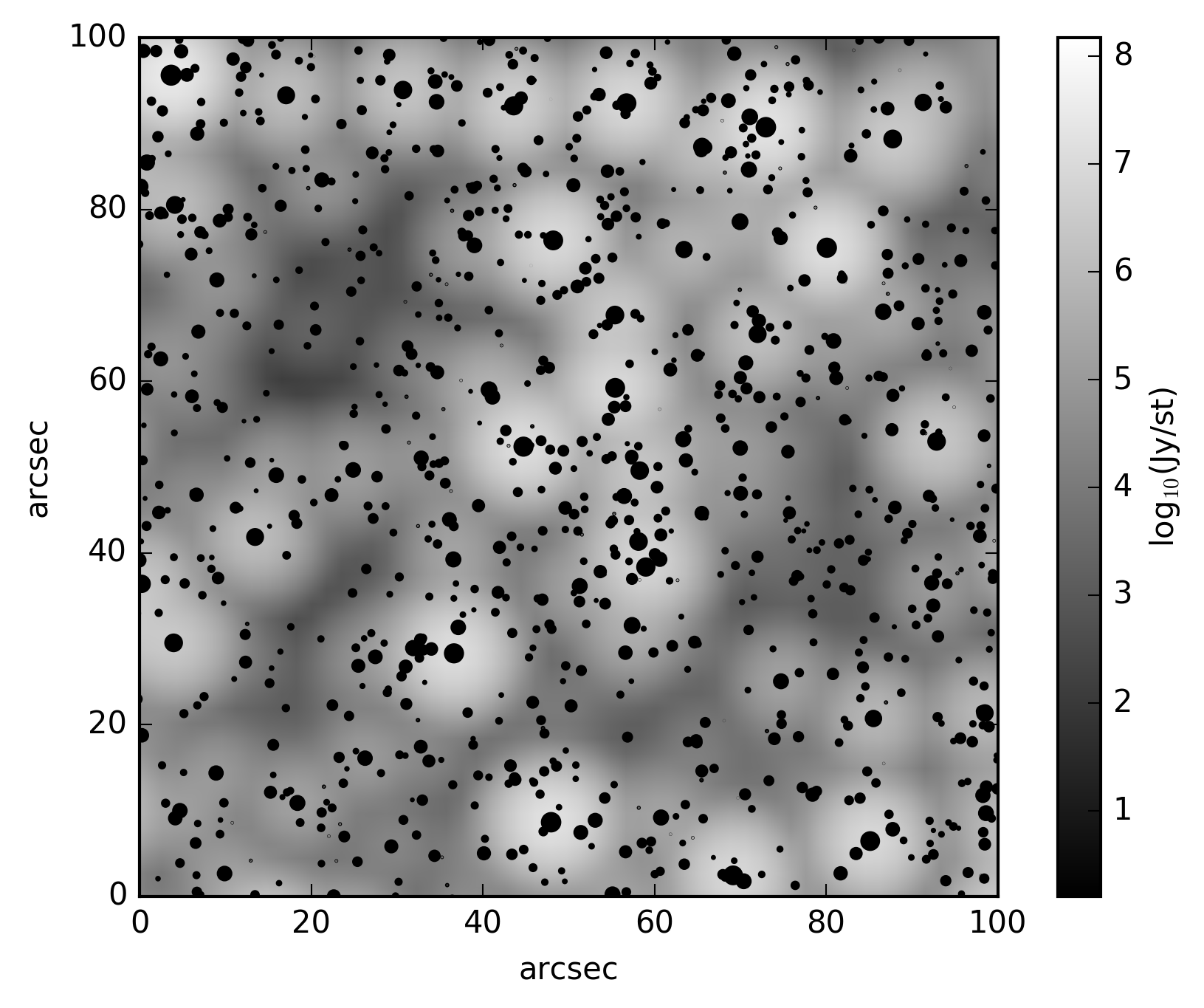}
\caption{Example of a $100''\times 100$ field of the simulated sky background map at 300 $\mu$m. The crosses on the left indicate
the intensity peaks found using the second derivatives. The dots on the right indicate the model galaxies. The dot size is
proportional to the logarithm of the galaxy flux.}
\label{fig:peaks}
\end{figure*}

Figure 3 illustrates the confusion problem, where
the distribution of model galaxies in the sky field
is shown on the right. After convolution with the
telescope beam, the brightest galaxies create aureoles
around themselves that exceed considerably $\lambda/D=6$'' in their sizes and in which the emission from the
neighboring fainter galaxies is lost. As a result, the
intensity peaks on the CIB map (in Fig. 3 on the left)
correspond to the brightest galaxies.

The results of determining the confusion limits
for two telescope apertures, 3.5 m (Herschel) and
10 m (Millimetron), for the four above-listed criteria
are presented in Table 3. It can be seen from this
table that the confusion limits for the 3.5-m telescope
predicted in our model according to the completeness
criterion are close to those actually found from the
Herschel observations. For example, the confusion
limit is 5.8 mJy for the SPIRE instrument at 250 $\mu$m
\citep{2010A&A...518L...5N} and 0.27 mJy for the PACS
instrument at 100 $\mu$m \citep{2013Tuttlebee}.

To determine the accuracy of finding the confusion
limit from the simulated maps, we divided the maps
into four equal regions and found the confusion limit
in each of them. The deviation from the limit for the
complete map turned out to be 60\% for all ranges and
apertures from Table 3.

\begin{table}
\caption{Confusion limits in the far-IR/submillimeter range}
\begin{center}
\begin{tabular}{lccc}
\hline
Aperture/criterion & 100 $\mu$m & 300 $\mu$m & 1000 $\mu$m  \\ \hline
3.5 m (Herschel) \\
Source density & 80 $\mu$Jy & 10 mJy & 3 mJy \\
Photometric (from counts) & 0 & 6 mJy & 5 mJy \\
Photometric (from map) & 60 $\mu$Jy & 20 mJy & 13 mJy \\
Completeness 50\%: &&& \\
\hspace{10pt}search for maxima & 0.2 mJy & 9 mJy & 4 mJy \\ 
\hspace{10pt}getsources monochromatic & 0.5 mJy & 14 mJy & 12 mJy \\
\hspace{10pt}getsources multiwavelength & 0.5 mJy & 3 mJy & 2 mJy \\ \hline

10 m (Millimetron) \\
Source density & 1 nJy & 0.3 mJy & 1 mJy \\
Photometric (from counts) & 0 & 0.3 $\mu$Jy & 1 mJy \\
Photometric (from map) & 40 $\mu$Jy & 0.3 mJy & 4 mJy \\
Completeness 50\%: &&& \\
\hspace{10pt}search for maxima & 14 $\mu$Jy & 0.6 mJy & 1 mJy \\ 
\hspace{10pt}getsources monochromatic & 9 $\mu$Jy & 0.7 mJy & 1 mJy \\
\hspace{10pt}getsources multiwavelength & 10 $\mu$Jy & 0.1 mJy & 1 mJy \\ 
\hline
\end{tabular}
\end{center}
\label{tab:conf}
\end{table}

We see that at 100 $\mu$m the photometric criterion
calculated from the source counts gives zero confusion limit. In addition, for the 10-m telescope at
300 $\mu$m the two photometric criteria give values differing by three orders of magnitude. This is because
the dependence $\sigma(S_\mathrm{lim})$ in some ranges of fluxes is
nearly linear, and a small shift leads to the disappearance of the solution of Eq. (8). The photometric
criterion uses only the tail of the distribution with the
lowest fluxes while disregarding the resolved sources
whose density in the sky is, nevertheless, quite high.
Therefore, in the case where the confusion limit according to the photometric criterion approaches zero,
other criteria should be used.

To check the influence of the large-scale structure
on confusion, we mixed the coordinates of galaxies
in the sky, leaving the fluxes from them unchanged.
Within the accuracy of our method no difference was
found for all the methods of determining the confusion
limit presented in Table 3.

\section{The Large-Scale Structure}

The CIB is known to be anisotropic. For example, a study of the spatial anisotropy power spectrum
obtained from the Herschel observations in \citet{2013ApJ...779...32V} showed that this spectrum could be
explained in terms of the analytical halo clustering
model. The constructed power spectrum for the simulated CIB maps obtained in our model for the 3.5-m
aperture is compared with the power spectrum from
\citet{2013ApJ...779...32V}. These spectra are shown in
Fig. 4. To construct them, we subtracted the shot-noise contribution from the power spectrum of the
spatial background anisotropy while taking into account the telescope beam.

\begin{figure*}
\includegraphics[width=\linewidth]{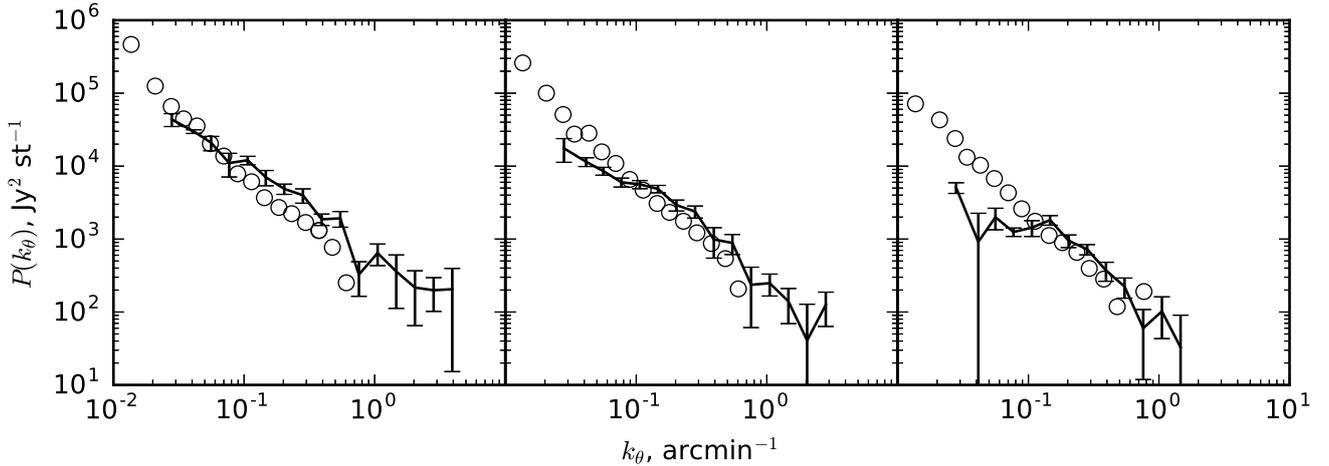}
\caption{Power spectrum of the spatial sky CIB anisotropy at 250, 350, and 500 $\mu$m (from left to right). The circles are the results from \citet{2013ApJ...779...32V} 
(no errors are shown); the solid line indicates our results.}
\end{figure*}

As can be seen from Fig. 4, our model success-
fully reproduces the background anisotropy spectrum
within the measurement error limits (the measurement errors at the Herschel telescope are not shown
in Fig. 4). The slight systematic differences may
stem from the fact that our map processing procedure
differs from that used in \citet{2013ApJ...779...32V}. In particular, to remove the shot noise, we produced the maps
without anisotropy where the source coordinates were
random, while their fluxes did not change compared to
the map with a structure.

It is important to note that in our model the elements of the three-dimensional large-scale structure
observed in the distribution of halos in space can be
directly compared with the clusters of spots on the
CIB map. For our cluster analysis we used the minimal spanning tree technique. We varied the threshold
length for three-dimensional clusters by choosing it
from the set of 0.3, 0.5, and 1.0 Mpc and the minimum
mass as 30, 100, or 500 objects, respectively. We
identified all peaks on the map with a brightness
twice as high as the confusion limits that we deter-
mined from the 50\% completeness criterion at 100
and 300 $\mu$m. Among these peaks we found clusters
with threshold branch lengths of 20'', 25'', and 30''
using the same minimal spanning tree algorithm; the
number of peaks is 10, 30, and 100, respectively.
Next, for all identifications of three-dimensional clusters and all identifications of cluster peaks we performed a cross-identification by their locations in the
sky. As a rule, one or more overlapping clusters were
observed. The probability of a chance coincidence
was determined by cyclically shifting one of the maps
by a random vector. As a result, it emerged that
the probability of a chance realization of the observed
coincidences at the parameters we use is at least 5\%,
suggesting that there is no correspondence between
the maps of peaks in the CIB and three-dimensional
superclusters of galaxies.

\section{Conclusions}

We constructed a semi-analytical model of the sky
CIB whose new element is allowance for the data from
cosmological simulations of the large-scale structure
of the Universe and the construction of simulated CIB
maps to which various source searching algorithms
can be applied. The developed model shows good
agreement with the known data on source counts and
the power spectrum of the spatial CIB anisotropy in
the wavelength range from 100 $\mu$m to 2 mm. This
model was used to determine the confusion limit for
future 10-m far-IR space telescopes and to compare
the clustering of background intensity peaks with the
actual large-scale structure.

Based on the source counts, we used the source
density criterion and the photometric criterion to estimate the confusion limit. The confusion limits obtained with these criteria were compared between
themselves and with the confusion limit obtained directly from the CIB map using the completeness criterion. In the wavelength range 300--1000 $\mu$m all of
the methods used showed close values of the confusion limit, while for a wavelength of 100 $\mu$m the results of the two estimation criteria differ significantly.
Furthermore, for the same wavelength, 100 $\mu$m, these
estimation criteria yielded a result significantly differing from that obtained by a direct measurement from
the background map. It is important to note that the
confusion limits for a 3.5-m telescope obtained with
the completeness criterion from the simulated map
turned out to be close to those found from the observations with the Herschel telescope. Hence it can be
concluded that at a telescope aperture of 3--10 m the
estimation criteria work well at comparatively long
wavelengths, $\lambda\geq 300$, and poorly at short ones.

At wavelengths of $\sim$100 $\mu$m it will be possible to
identify compact sources with a flux density above
10 $\mu$Jy from the maps obtained in the mode of broad-band photometry with a 10-m telescope; this is better
than that for a telescope with a mirror diameter of
3.5 m by more than an order of magnitude. For a
wavelength of 300 $\mu$m, at which the CIB intensity
is at a maximum, the confusion limit will be about
0.6 mJy, which is also lower than the confusion limit
measured with the Herschel telescope by an order
of magnitude. For a wavelength of 1 mm a 10-m
telescope gives a fourfold gain in confusion limit,
reaching 1 mJy.

The CIB in our model demonstrates significant
deviations from a uniform random distribution of
sources in the sky. In this case, the constructed
power spectrum of the spatial background anisotropy
within the developed model shows good agreement
with the power spectrum obtained from the Herschel
observations. Thus, the large-scale structure of the
Universe clearly manifests itself on the CIB maps.
At the same time, the large-scale structure has no
noticeable influence on the confusion limit found from
the simulated maps.

Our model has demonstrated for the first time that
the fluctuations in the number of intensity peaks ob-
served in a 1$^\circ\times1^\circ$ field show no clear correlation with
superclusters of galaxies. It is necessary to invoke
spectroscopic information, for example, to use [CII]
and CO lines, to identify the three-dimensional large-scale structure.

Various ways of breaking the confusion limit are
considered in the literature: \cite{Raymond10}
assessed the true viability of using spectroscopic information to reduce the confusion on the planned
Japanese SPICA space telescope. \cite{Safarzadeh14} studied the possibility of using optical information to reduce the confusion in the far IR on Herschel images. It is necessary to carry out such studies
for Millimetron and other planned observatories, for
which purpose our developed model is planned to be
used.

\section{Acknowledgements}

The CosmoSim database used in this paper is
a service provided by the Leibnitz Institute for Astrophysics Potsdam. We thank the Gauss Centre for Supercomputing (www.gauss-centre.eu) and
the Partnership for Advanced Computing in Europe
(PRACE, www.prace-ri.eu) for supporting the MultiDark project of numerical simulations and for providing the computational time for it at GCS Supercomputer SuperMUC at the Leibnitz Rechenzentrum
(LRZ, www.lrz.de).

The work of S.V. Pilipenko, M.V. Tkachev,
A.A. Ermash, E.V. Mikheeva, and V.N. Lukash
was supported by the Russian Foundation for Basic
Research grant no. 16-02-01043. The work of
M.V. Tkachev was also supported by the Russian
Foundation for Basic Research grant no. 16-32-00263. The work was supported by the Basic
Research Program P-7 of the Presidium of the
Russian Academy of Sciences and grant no. NSh-6595.2016.2 from the President of the Russian Federation for Support of Leading Scientific Schools.

\bibliographystyle{mnras}
\bibliography{sky}
\bsp	
\label{lastpage}
\end{document}